\documentstyle[12pt]{article}

\begin{document}

\begin{titlepage}

\begin{center}

{\Large \bf
Lagrangian Gauge Structure Functions for Systems with First-Class
Constraints}
\vskip .6in

Domingo J. Louis-Martinez
\vskip .2in

Science One Program and\\ 
Department of Physics and Astronomy,\\ University of British
Columbia\\Vancouver, Canada

\end{center}
 
\vskip 3cm
 
\begin{abstract}
The structure functions of the lagrangian gauge algebra are given
explicitly in terms of the hamiltonian constraints and the first-order
hamiltonian structure functions and their derivatives. 
\end{abstract} 
\end{titlepage}

The importance of gauge symmetries in all modern relativistic theories of
the fundamental interactions is well understood at present.
Electromagnetic and Yang-Mills theories are examples of systems
with natural bases of gauge generators that form closed lagrangian gauge
algebras. The hamiltonian first-class constraints in these theories are
linear in the canonical momenta. Their first-order
hamiltonian gauge structure functions are constants and therefore the
first-class constraints form Lie groups. In Einstein's theory of gravity
one of the hamiltonian constraints is quadratic in the momenta
\cite{dirac1} and the first-order hamiltonian gauge structure functions
do not depend on the canonical momenta but do depend on the metric
\cite{fradkin}. The hamiltonian constraints in Einstein's theory do not
form a Lie group \cite{fradkin}, but the natural basis of
lagrangian gauge generators does form a closed gauge algebra.

The study of open lagrangian gauge algebras started with the discovery of
supergravity \cite{freedman, kallosh, dewit}. In theories with open 
lagrangian gauge algebras we have to deal with lagrangian structure
functions of higher orders. The
existence of these structure functions has been established using an
axiomatic approach in \cite{batalin2}. In the quantum theory, in order to
construct the Feynman diagrams one needs first to determine all the gauge
structure functions. Although these functions have been
found explicitly in particular cases \cite{kallosh}, their general
form for the generic case of a hamiltonian system with first-class
constraints is not known.
The purpose of this paper is to solve this problem. We will present
explicit expressions for the lagrangian gauge structure functions up to
fourth order. 
We will show  how these structure tensors 
are determined by the hamiltonian constraints and the hamiltonian
first-order structure functions. It is remarkable that to determine the
higher order lagrangian gauge structure tensors no knowledge of the
higher-order hamiltonian structure functions is required. They depend only
on the zeroth- and first-order hamiltonian structure functions.
The method presented here can be used to find the structure functions of
even higher levels.

Let us consider a physical system described by the action:

\begin{equation}
S = \int dt L
\label{1.1}
\end{equation} 

$L$ is the Lagrangian of the system, which is a function defined in the
velocity phase space $TQ$ ($TQ$ is the tangent bundle of the n-dimensional
configuration space $Q$). The variables $q^{i}$ ($i=1,2,...,n$) are the
generalized coordinates and $\dot{q}^{i}$ the generalized velocities.

The Euler-Lagrange equations of motion may be written in the following
form:

\begin{equation}
L_{i} \equiv W_{ij}\ddot{q}^{j} - \alpha_{i} = 0
\label{1.2}
\end{equation}

\noindent where, 

\begin{equation}
W_{ij} \equiv 
\frac{\partial^{2} L}{\partial\dot{q}^{i} \partial\dot{q}^{j}}
\label{1.3}
\end{equation}

\begin{equation}
\alpha_{i} \equiv \frac{\partial L}{\partial q^{i}} - 
\dot{q}^{l} \frac{\partial^{2} L}{\partial q^{l} \partial\dot{q}^{i}}
\label{1.4}
\end{equation} 

We assume that the dimension of the kernel of the Hessian matrix $W$ is
constant (the same in all points of $TQ$) and equal to $m$:

\begin{equation}
rank\| W \| = n-m
\label{1.5}
\end{equation}

Let us denote as $R_{\mu}(q,\dot{q})$ ($\mu = 1,2,...,m$) a set of
linearly independent null eigenvectors of the Hessian matrix:

\begin{equation}
R_{\mu}^{i} W_{ij} \equiv 0
\label{1.6}
\end{equation}

\begin{equation}
rank \| R_{\mu}^{i} \| = m
\label{1.7}
\end{equation}

Using the Euler-Lagrange equations (\ref{1.2}) and the identities
(\ref{1.6}) we obtain the lagrangian constraints of the first-level
\cite{sudarshan}:

\begin{equation}
\chi^{(1)}_{\mu} \equiv R_{\mu}^{i}\alpha_{i}=0
\label{1.8}
\end{equation}

Let us assume that (\ref{1.8}) do not bring about any restrictions in
the velocity phase space $TQ$ (we assume that (\ref{1.8}) are identities
in $TQ$):

\begin{equation}
R_{\mu}^{i} \alpha_{i} \equiv 0
\label{1.9}
\end{equation}

In this case, the left-hand-sides $L_{i}$ of the Euler-Lagrange equations 
(\ref{1.2}) satisfy the Noether identities \cite{sudarshan}:

\begin{equation}
R_{\mu}^{i}(q,\dot{q}) L_{i}(q,\dot{q},\ddot{q}) \equiv 0
\label{1.10}
\end{equation}

The Noether identities are satisfied by any trajectory $q(t)$ in the
configuration space Q. According to the second Noether theorem 
\cite{gitman}, the action (\ref{1.1}) is invariant under the
infinitesimal gauge transformations:

\begin{equation}
\delta q^{i}(t) = \epsilon^{\mu}(t) R_{\mu}^{i}(q(t),\dot{q}(t))
\label{1.11}
\end{equation}

\noindent $R_{\mu}^{i}$ are the generators of the gauge transformations
(\ref{1.11}).

It is not difficult to prove that under infinitesimal gauge
transformations of the form (\ref{1.11}) the left-hand-sides of the
Euler-Lagrange equations (\ref{1.2}) transform as follows:

\begin{equation}
\delta L_{i} = \epsilon^{\mu} \left[ - \frac{\partial
R_{\mu}^{j}}{\partial
q^{i}} L_{j} + \frac{d}{dt} (\frac{\partial
R_{\mu}^{j}}{\partial\dot{q}^{i}} L_{j} )\right] + \dot{\epsilon}^{\mu}
\frac{\partial R_{\mu}^{j}}{\partial\dot{q}^{i}} L_{j}
\label{1.13}
\end{equation}

From (\ref{1.7}) it follows that we are dealing with an irreducible gauge
theory \cite{dewit, batalin1}. The most general regular solution of
the equation:

\begin{equation}
\lambda^{i} L_{i} \equiv 0
\label{1.14}
\end{equation}

\noindent can be written as \cite{batalin2}:

\begin{equation}
\lambda^{i} = R_{\mu}^{i} \epsilon^{\mu} + T^{ij} L_{j}
\label{1.15}
\end{equation}

\noindent where,

\begin{equation}
T^{ij} = - T^{ji}
\label{1.16}
\end{equation}

The trivial gauge transformations $T^{ij} L_{j}$ have no physical
significance. However, in relativistic field theories trivial gauge
transformations may appear as a result of requiring manifest covariance and
preservation of locality.

Let us consider the commutator of two infinitesimal gauge transformations:

\begin{equation}
\delta_{1} \delta_{2} q^{i}  - \delta_{2} \delta_{1} q^{i} \equiv
[\delta_{1},\delta_{2}] q^{i}  
\label{1.17}
\end{equation}

In general, the commutator of two infinitesimal gauge transformations of
the form (\ref{1.11}) is a gauge transformation of the form (\ref{1.15})
\cite{dewit, batalin1, batalin2}:

\begin{equation}
[\delta_{1},\delta_{2}] q^{i} \equiv R_{\gamma}^{i} \epsilon^{\gamma} +
E^{ij} L_{j} 
\label{1.18}
\end{equation}

\noindent where,

\begin{equation}
\epsilon^{\gamma} = T_{\alpha\beta}^{\gamma}(q, \dot{q})
\epsilon_{1}^{\alpha} \epsilon_{2}^{\beta}
\label{1.19}
\end{equation}

\begin{equation}
T_{\alpha\beta}^{\gamma} = - T_{\beta\alpha}^{\gamma}
\label{1.20}
\end{equation}

\begin{equation}
E^{ij} = E_{\alpha\beta}^{ij}(q, \dot{q}) \epsilon_{1}^{\alpha}
\epsilon_{2}^{\beta}
\label{1.21}
\end{equation}

\begin{equation}
E_{\alpha\beta}^{ij} = - E_{\alpha\beta}^{ji} = - E_{\beta\alpha}^{ij}
\label{1.22}
\end{equation}

The lagrangian gauge generators $R_{\mu}^{i}$ form an open gauge algebra 
if $E_{\alpha\beta}^{ij} \neq 0$ \cite{gitman, dewit, batalin1, batalin2,
gomis}. If $E_{\alpha\beta}^{ij} = 0$, the gauge algebra is said to be
closed\footnote{Notice that by adding trivial gauge transformations to
the non-trivial gauge transformations (\ref{1.11}) a closed gauge algebra
can easily be converted into an "open" gauge algebra. However, in this
case the algebra is not really open but "hidden closed". I would like to
thank the referee for pointing this out.}.

The tensors $T_{\alpha\beta}^{\gamma}$ and $E_{\alpha\beta}^{ij}$ are the
so-called
second-order gauge structure functions of the gauge algebra in the
lagrangian formalism. The gauge generators $R_{\mu}^{i}$ are the
first-order structure functions of the lagrangian gauge algebra. The
zeroth-order structure function is the action S itself \cite{batalin2}.

Since $\epsilon_{1}^{\mu}$ and $\epsilon_{2}^{\nu}$ are
arbitrary functions of time, then from (\ref{1.18}) we obtain the
following identities in the velocity phase space $TQ$:

\begin{equation}
\frac{\partial R_{\mu}^{i}}{\partial q^{j}} R_{\nu}^{j} - 
\frac{\partial R_{\nu}^{i}}{\partial q^{j}} R_{\mu}^{j} 
+ \left(\frac{\partial R_{\mu}^{i}}{\partial\dot{q}^{j}} 
\frac{\partial R_{\nu}^{j}}{\partial q^{k}} - 
\frac{\partial R_{\nu}^{i}}{\partial\dot{q}^{j}} 
\frac{\partial R_{\mu}^{j}}{\partial q^{k}}\right) \dot{q}^{k}
\equiv T_{\mu\nu}^{\gamma} R_{\gamma}^{i} - E_{\mu\nu}^{ij} \alpha_{j}
\label{1.23}
\end{equation}

\begin{equation}
\frac{\partial R_{\mu}^{i}}{\partial\dot{q}^{j}} 
\frac{\partial R_{\nu}^{j}}{\partial\dot{q}^{k}} - 
\frac{\partial R_{\nu}^{i}}{\partial\dot{q}^{j}} 
\frac{\partial R_{\mu}^{j}}{\partial\dot{q}^{k}} 
\equiv E_{\mu\nu}^{ij} W_{jk}
\label{1.24}
\end{equation}

\begin{equation}
\frac{\partial R_{\mu}^{i}}{\partial \dot{q}^{j}} R_{\nu}^{j} \equiv 0
\label{1.25}
\end{equation}

It is not difficult to prove that the identities (\ref{1.23}-\ref{1.25})
are equivalent to the so-called second-order relations of the lagrangian
gauge algebra given in \cite{batalin2, gomis}.

From (\ref{1.5},\ref{1.6}) and (\ref{1.25}) it follows \cite{masud} that
there exist some functions $b_{\mu}^{ij}(q, \dot{q})$ such that:

\begin{equation}
\frac{\partial R_{\mu}^{i}}{\partial\dot{q}^{j}} \equiv b_{\mu}^{ik}
W_{kj}
\label{1.26}
\end{equation}

Substituting (\ref{1.26}) into (\ref{1.24}) we find the following
identities:

\begin{equation}
(E_{\mu\nu}^{ij} - b_{\mu}^{im} W_{mn} b_{\nu}^{nj}
+ b_{\nu}^{im} W_{mn} b_{\mu}^{nj}) W_{jk} \equiv 0
\label{1.27}
\end{equation}

From (\ref{1.6}), (\ref{1.9}), (\ref{1.23}) and (\ref{1.24}) one can see
that the structure functions $E_{\mu\nu}^{ij}$ are not uniquely
determined. In general, one can write:

\begin{equation}
E_{\mu\nu}^{ij} = b_{\mu}^{im} W_{mn} b_{\nu}^{nj}
- b_{\nu}^{im} W_{mn} b_{\mu}^{nj} + 
e_{\mu\nu}^{\alpha\beta}(R_{\alpha}^{i} R_{\beta}^{j} - R_{\beta}^{i}
R_{\alpha}^{j}) 
\label{1.28}
\end{equation}

\noindent where $e_{\mu\nu}^{\alpha\beta}(q, \dot{q})$ are arbitrary
functions on $TQ$ that are antisymmetric in the lower indexes:

\begin{equation}
e_{\mu\nu}^{\alpha\beta} = - e_{\nu\mu}^{\alpha\beta}
\label{1.29}
\end{equation}

From (\ref{1.26}) and (\ref{1.28}) it immediately follows that:

\begin{equation}
\frac{\partial R_{\alpha}^{i}}{\partial \dot{q}^{k}} E_{\beta\gamma}^{kj}
+ \frac{\partial R_{\beta}^{i}}{\partial \dot{q}^{k}}
E_{\gamma\alpha}^{kj}
+ \frac{\partial R_{\gamma}^{i}}{\partial \dot{q}^{k}}
E_{\alpha\beta}^{kj}
+ \frac{\partial R_{\alpha}^{j}}{\partial \dot{q}^{k}}
E_{\beta\gamma}^{ki}
+ \frac{\partial R_{\beta}^{j}}{\partial \dot{q}^{k}}
E_{\gamma\alpha}^{ki}
+ \frac{\partial R_{\gamma}^{j}}{\partial \dot{q}^{k}}
E_{\alpha\beta}^{ki}
\equiv 0
\label{1.30}
\end{equation}

The third-order relations of the lagrangian gauge algebra can be derived
from the Jacobi identities \cite{dewit, batalin1, batalin2, gomis}:

\begin{equation}
[\delta_{1},[\delta_{2},\delta_{3}]] + 
[\delta_{2},[\delta_{3},\delta_{1}]] + 
[\delta_{3},[\delta_{1},\delta_{2}]]
\equiv 0
\label{1.31}
\end{equation}

From (\ref{1.31}), using (\ref{1.11}, \ref{1.13}, \ref{1.18}-\ref{1.25})
and (\ref{1.30}), we obtain the following identities:

\begin{equation}
R_{\rho}^{i} A_{\alpha\beta\gamma}^{\rho} + B_{\alpha\beta\gamma}^{ij}
L_{j} \equiv 0
\label{1.37}
\end{equation}

\begin{equation}
- R_{\eta}^{i} R_{\alpha}^{k}
\frac{\partial T_{\beta\gamma}^{\eta}}{\partial\dot{q}^{k}}
- \frac{\partial R_{\eta}^{i}}{\partial\dot{q}^{k}} R_{\alpha}^{k}
T_{\beta\gamma}^{\eta}
+ \left( \frac{\partial R_{\beta}^{i}}{\partial\dot{q}^{k}}
E_{\gamma\alpha}^{kj}
+ \frac{\partial R_{\gamma}^{i}}{\partial\dot{q}^{k}} E_{\alpha\beta}^{kj}
+ \frac{\partial R_{\alpha}^{j}}{\partial\dot{q}^{k}} E_{\beta\gamma}^{ki}
- R_{\alpha}^{k}
\frac{\partial E_{\beta\gamma}^{ij}}{\partial\dot{q}^{k}}\right)
L_{j} \equiv 0
\label{1.35}
\end{equation}

\noindent where,

\begin{eqnarray}
A_{\alpha\beta\gamma}^{\rho} 
&\equiv& {1 \over 3} \left[T_{\alpha\eta}^{\rho} T_{\beta\gamma}^{\eta} 
+ T_{\beta\eta}^{\rho} T_{\gamma\alpha}^{\eta} 
+ T_{\gamma\eta}^{\rho} T_{\alpha\beta}^{\eta}\right. \nonumber \\
& & - R_{\alpha}^{j} \frac{\partial T_{\beta\gamma}^{\rho}}{\partial q^{j}}  
- R_{\beta}^{j} \frac{\partial T_{\gamma\alpha}^{\rho}}{\partial q^{j}}  
- R_{\gamma}^{j} \frac{\partial T_{\alpha\beta}^{\rho}}{\partial q^{j}}
\nonumber \\
& & 
\left.-\dot{R}_{\alpha}^{j} \frac{\partial
T_{\beta\gamma}^{\rho}}{\partial\dot{q}^{j}}  
-\dot{R}_{\beta}^{j} \frac{\partial T_{\gamma\alpha}^{\rho}}{\partial
\dot{q}^{j}}  
-\dot{R}_{\gamma}^{j} \frac{\partial T_{\alpha\beta}^{\rho}}{\partial
\dot{q}^{j}}\right]
\label{1.38}
\end{eqnarray}

\begin{eqnarray}
B_{\alpha\beta\gamma}^{ij} &\equiv&
{1 \over 3}\left[ 
E_{\alpha\eta}^{ij} T_{\beta\gamma}^{\eta}
+ E_{\beta\eta}^{ij} T_{\gamma\alpha}^{\eta}
+ E_{\gamma\eta}^{ij} T_{\alpha\beta}^{\eta}\right. \nonumber \\
& &- R_{\alpha}^{k} \frac{\partial E_{\beta\gamma}^{ij}}{\partial q^{k}}
- R_{\beta}^{k} \frac{\partial E_{\gamma\alpha}^{ij}}{\partial q^{k}}
- R_{\gamma}^{k} \frac{\partial E_{\alpha\beta}^{ij}}{\partial q^{k}}
- \dot{R}_{\alpha}^{k} \frac{\partial E_{\beta\gamma}^{ij}}{\partial
\dot{q}^{k}}
- \dot{R}_{\beta}^{k} \frac{\partial E_{\gamma\alpha}^{ij}}{\partial
\dot{q}^{k}}
- \dot{R}_{\gamma}^{k} \frac{\partial E_{\alpha\beta}^{ij}}{\partial
\dot{q}^{k}} \nonumber \\
& & + \frac{\partial R_{\alpha}^{i}}{\partial q^{k}} E_{\beta\gamma}^{kj} 
+ \frac{\partial R_{\beta}^{i}}{\partial q^{k}} E_{\gamma\alpha}^{kj} 
+ \frac{\partial R_{\gamma}^{i}}{\partial q^{k}} E_{\alpha\beta}^{kj} 
- \frac{\partial R_{\alpha}^{j}}{\partial q^{k}} E_{\beta\gamma}^{ki} 
- \frac{\partial R_{\beta}^{j}}{\partial q^{k}} E_{\gamma\alpha}^{ki} 
- \frac{\partial R_{\gamma}^{j}}{\partial q^{k}} E_{\alpha\beta}^{ki}
\nonumber \\ 
& & + \frac{\partial R_{\alpha}^{i}}{\partial \dot{q}^{k}}
\dot{E}_{\beta\gamma}^{kj} 
+ \frac{\partial R_{\beta}^{i}}{\partial \dot{q}^{k}}
\dot{E}_{\gamma\alpha}^{kj} 
+ \frac{\partial R_{\gamma}^{i}}{\partial \dot{q}^{k}}
\dot{E}_{\alpha\beta}^{kj} 
- \frac{\partial R_{\alpha}^{j}}{\partial \dot{q}^{k}}
\dot{E}_{\beta\gamma}^{ki} 
- \frac{\partial R_{\beta}^{j}}{\partial \dot{q}^{k}}
\dot{E}_{\gamma\alpha}^{ki} 
- \frac{\partial R_{\gamma}^{j}}{\partial \dot{q}^{k}}
\dot{E}_{\alpha\beta}^{ki} \nonumber\\
& &
+ {1 \over 2} \left.\frac{d}{dt} \left(
\frac{\partial R_{\alpha}^{j}}{\partial\dot{q}^{k}} E_{\beta\gamma}^{ki}
+ \frac{\partial R_{\beta}^{j}}{\partial\dot{q}^{k}} E_{\gamma\alpha}^{ki}
+ \frac{\partial R_{\gamma}^{j}}{\partial\dot{q}^{k}} E_{\alpha\beta}^{ki}
- \frac{\partial R_{\alpha}^{i}}{\partial\dot{q}^{k}} E_{\beta\gamma}^{kj}
- \frac{\partial R_{\beta}^{i}}{\partial\dot{q}^{k}} E_{\gamma\alpha}^{kj}
- \frac{\partial R_{\gamma}^{i}}{\partial\dot{q}^{k}}
E_{\alpha\beta}^{kj}\right)\right] \nonumber \\
\label{1.39}
\end{eqnarray}

Since, by assumption, the gauge generators $R_{\rho}^{i}$ are irreducible,
then from (\ref{1.37}) it follows that there must exist some functions
$D_{\alpha\beta\gamma}^{i\rho}$ such that \cite{batalin2, gomis}:

\begin{equation}
A_{\alpha\beta\gamma}^{\rho} \equiv D_{\alpha\beta\gamma}^{i\rho} L_{i}
\label{1.41}
\end{equation}

The identities (\ref{1.41}) are the so-called third-order relations of the
lagrangian gauge algebra \cite{batalin2}. $D_{\alpha\beta\gamma}^{i\rho}$
are the third-order gauge structure functions in the lagrangian formalism. 

From (\ref{1.41}) we see that the third-order structure functions are not
uniquely determined. Indeed, if 
$D_{\alpha\beta\gamma}^{i\rho}(q,\dot{q})$ satisfy (\ref{1.41}), then
using (\ref{1.38}) and (\ref{1.2}) we can rewrite (\ref{1.41}) as
identities in $TQ$:

\begin{eqnarray}
{1 \over 3} \left[T_{\alpha\eta}^{\rho} T_{\beta\gamma}^{\eta} 
+ T_{\beta\eta}^{\rho} T_{\gamma\alpha}^{\eta} 
+ T_{\gamma\eta}^{\rho} T_{\alpha\beta}^{\eta}
- R_{\alpha}^{j} \frac{\partial T_{\beta\gamma}^{\rho}}{\partial q^{j}}  
- R_{\beta}^{j} \frac{\partial T_{\gamma\alpha}^{\rho}}{\partial q^{j}}  
- R_{\gamma}^{j} \frac{\partial T_{\alpha\beta}^{\rho}}{\partial
q^{j}}\right. & &\nonumber \\
\left.- \left(\frac{\partial R_{\alpha}^{j}}{\partial q^{l}}
\frac{\partial T_{\beta\gamma}^{\rho}}{\partial\dot{q}^{j}}  
+ \frac{\partial R_{\beta}^{j}}{\partial q^{l}} 
\frac{\partial T_{\gamma\alpha}^{\rho}}{\partial\dot{q}^{j}}  
+ \frac{\partial R_{\gamma}^{j}}{\partial q^{l}} 
\frac{\partial T_{\alpha\beta}^{\rho}}{\partial\dot{q}^{j}}\right) 
\dot{q}^{l}\right] &\equiv& - D_{\alpha\beta\gamma}^{i\rho} \alpha_{i}
\label{1.381}
\end{eqnarray}

\begin{eqnarray}
- {1 \over 3} \left[\frac{\partial R_{\alpha}^{j}}{\partial \dot{q}^{k}}
\frac{\partial T_{\beta\gamma}^{\rho}}{\partial\dot{q}^{j}}  
+ \frac{\partial R_{\beta}^{j}}{\partial \dot{q}^{k}} 
\frac{\partial T_{\gamma\alpha}^{\rho}}{\partial\dot{q}^{j}}  
+ \frac{\partial R_{\gamma}^{j}}{\partial \dot{q}^{k}} 
\frac{\partial T_{\alpha\beta}^{\rho}}{\partial\dot{q}^{j}}\right] 
\equiv  D_{\alpha\beta\gamma}^{i\rho} W_{ik}
\label{1.382}
\end{eqnarray}

From (\ref{1.381}, \ref{1.382}) and (\ref{1.6}, \ref{1.9}) we conclude
that any function $\tilde{D}_{\alpha\beta\gamma}^{i\rho}$:

\begin{equation}
\tilde{D}_{\alpha\beta\gamma}^{i\rho} = D_{\alpha\beta\gamma}^{i\rho} +
d_{\alpha\beta\gamma}^{\rho\delta} R_{\delta}^{i}
\label{1.42}
\end{equation} 

\noindent must also be a solution of (\ref{1.381}, \ref{1.382}).
The quantities $d_{\alpha\beta\gamma}^{\rho\delta}$ are arbitrary
functions.

Finally, from (\ref{1.37}) and (\ref{1.41}), using the
properties of irreducibility (\ref{1.7}) and completeness
(\ref{1.14},\ref{1.15}), one can obtain the fourth-order relations 
of the lagrangian gauge algebra in the form \cite{gomis}:

\begin{equation}
R_{\rho}^{i} D_{\alpha\beta\gamma}^{j\rho} 
- R_{\rho}^{j} D_{\alpha\beta\gamma}^{i\rho}
+ B_{\alpha\beta\gamma}^{ij}
\equiv M_{\alpha\beta\gamma}^{ijk} L_{k}
\label{1.45}
\end{equation}

The tensors $M_{\alpha\beta\gamma}^{ijk}$ are the fourth-order gauge
structure functions in the lagrangian formalism.

Higher order gauge structure relations can be obtained by commuting a
higher number of infinitesimal gauge transformations. 
All the lagrangian gauge structure relations are encoded in the
so-called classical master equation obeyed by the field-antifield action
functional \cite{batalin2, gomis}. 
The lagrangian gauge algebra is characterized by the whole set of gauge
structure functions.

The existence of the lagrangian gauge structure functions
$D_{\alpha\beta\gamma}^{i\rho}$ and $M_{\alpha\beta\gamma}^{ijk}$ 
has been proven using an axiomatic approach \cite{batalin2, gomis}. 
In this paper we aim to construct these functions explicitly.

With this purpose in mind, let us develop the hamiltonian formulation for
the action functional (\ref{1.1}). From (\ref{1.5}) it follows that our
system has $m$ independent primary constraints $G_{\mu}$
$(\mu=1,2,...,m)$.  Following Dirac's method \cite{dirac}, these primary
constraints are obtained from the relations that define the canonical
momenta:

\begin{equation}
p_{i} = \frac{\partial L}{\partial\dot{q}^{i}}(q,\dot{q})
\label{2.1}
\end{equation}

From (\ref{2.1}) and (\ref{1.5}) we see that:

\begin{equation}
rank\|\frac{\partial G_{\mu}}{\partial p_{i}}\| = m
\label{2.2}
\end{equation}

The set of hamiltonian constraints:

\begin{equation}
G_{\mu}(q,p) = 0
\label{2.3}
\end{equation}

\noindent defines a submanifold of the momentum phase space $T^{*}Q$ which
is denoted by $\cal{M}$. 

The canonical Hamiltonian $H_{c}(q,p)$ is any function in the momentum
phase space $T^{*}Q$ satisfying the following relation:

\begin{equation}
{\rm FL}^* H_{c} \equiv H_{c}
\left(q, \frac{\partial L}{\partial\dot{q}}(q,\dot{q})\right) = 
\dot{q}^{i} \frac{\partial L}{\partial\dot{q}^{i}} - L
\label{2.7}
\end{equation}

Following \cite{battle} ${\rm FL}$ is the application fiber derivative of
the Lagrangian of the tangent bundle $TQ$ on the cotangent bundle $T^*Q$,

\[ {\rm FL}: \; TQ \longrightarrow T^*Q \]

\noindent given by ${\rm FL}(q,\dot{q}) = (q,p)$ as defined by
(\ref{2.1}). ${\rm FL}^*$ is the pullback application.

For the primary hamiltonian constraints $G_{\mu}$ we have:

\begin{equation}
{\rm FL}^* G_{\mu} = G_{\mu}\left(q, \frac{\partial
L}{\partial\dot{q}}(q,\dot{q})\right) \equiv 0
\label{2.8}
\end{equation}

Differentiating (\ref{2.8}) with respect to $\dot{q}^{j}$ we obtain:

\begin{equation}
{\rm FL}^* \frac{\partial G_{\mu}}{\partial p_{i}} W_{ij} \equiv 0 
\label{2.9}
\end{equation}

Equations (\ref{2.9}), (\ref{2.2}), (\ref{1.6}) and (\ref{1.7}) allow us
to identify ${\rm FL}^* \frac{\partial G_{\mu}}{\partial p_{i}}$ with the
irreducible gauge generators $R_{\mu}^{i}$:

\begin{equation}
R_{\mu}^{i}(q,\dot{q}) = 
{\rm FL}^* \frac{\partial G_{\mu}}{\partial p_{i}}
\label{2.10}
\end{equation}

\noindent In other words, one can always choose the constraints $G_{\mu}$
in such a way that the relations (\ref{2.10}) are true.

From (\ref{2.10}) and (\ref{2.8}) it also follows that:

\begin{equation}
- R_{\mu}^{j} \frac{\partial^{2} L}{\partial q^{i} \partial\dot{q}^{j}}
= {\rm FL}^* \frac{\partial G_{\mu}}{\partial q^{i}}
\label{2.11}
\end{equation}

From the definition of the canonical Hamiltonian (\ref{2.7}) we can also
derive the following relations \cite{masud}:

\begin{equation}
{\rm FL}^* \frac{\partial H_{c}}{\partial p_{i}} =
\dot{q}^{i} - \lambda_{\mu}(q,\dot{q}) R_{\mu}^{i}(q,\dot{q}) 
\label{2.12}
\end{equation}

\begin{equation}
{\rm FL}^* \frac{\partial H_{c}}{\partial q^{i}} =
- \frac{\partial L}{\partial q^{i}} + \lambda_{\mu}(q,\dot{q}) 
R_{\mu}^{j}(q,\dot{q}) 
\frac{\partial^{2} L}{\partial\dot{q}^{j}\partial q^{i}}
\label{2.13}
\end{equation}

Let us consider the Poisson brackets among the primary hamiltonian
constraints. From (\ref{2.10}) and (\ref{2.11}) we obtain:

\begin{equation}
{\rm FL}^* \{G_{\mu}, G_{\nu}\} 
\equiv - R_{\mu}^{i} B_{ij} R_{\nu}^{j}
\label{2.15}
\end{equation}

\noindent where,

\begin{equation}
B_{ij} = \frac{\partial^{2} L}{\partial\dot{q}^{i} \partial q^{j}} -
\frac{\partial^{2} L}{\partial\dot{q}^{j} \partial q^{i}}
\label{2.16}
\end{equation}

For the Poisson brackets between the constraints and the canonical
Hamiltonian we have:

\begin{equation}
{\rm FL}^* \{H_{c}, G_{\mu}\} \equiv
- R_{\mu}^{i} \alpha_{i} - R_{\mu}^{i} B_{ij} R_{\nu}^{j} \lambda_{\nu}
\label{2.17}
\end{equation}

Differentiating (\ref{1.9}) with respect to $\dot{q}^{j}$ and using
the identities (\ref{1.6}) we obtain the following identities in $TQ$:

\begin{equation}
\frac{\partial R_{\alpha}^{i}}{\partial\dot{q}^{j}} \alpha_{i}
- R_{\alpha}^{i} B_{ij} + \dot{q}^{l} \frac{\partial
R_{\alpha}^{i}}{\partial q^{l}} W_{ij} \equiv 0
\label{2.20}
\end{equation}

Multiplying (\ref{2.20}) by $R_{\beta}^{j}$ and using (\ref{1.25}) and
(\ref{1.6}) we obtain the identities:

\begin{equation}
R_{\alpha}^{i} B_{ij} R_{\beta}^{j} \equiv 0
\label{2.21}
\end{equation}

Therefore, from (\ref{2.15}), (\ref{2.17}), (\ref{1.9}) and (\ref{2.21})
we obtain the following identities in the velocity phase space $TQ$:

\begin{equation}
{\rm FL}^* \{G_{\mu}, G_{\nu}\} \equiv 0
\label{2.22}
\end{equation}

\begin{equation}
{\rm FL}^* \{H_{c}, G_{\mu}\} \equiv 0
\label{2.23}
\end{equation}

\noindent or equivalently, in the cotangent manifold $T^*Q$:

\begin{equation}
\{G_{\mu}, G_{\nu}\} \equiv C_{\mu\nu}^{\alpha} G_{\alpha}
\label{2.24}
\end{equation}

\begin{equation}
\{H_{c}, G_{\mu}\} \equiv V_{\mu}^{\alpha} G_{\alpha}
\label{2.25}
\end{equation}

\noindent where $C_{\mu\nu}^{\alpha}(q, p)$ and $V_{\mu}^{\alpha}(q, p)$
are funtions in $T^*Q$.

From the identities (\ref{2.24}) we conclude that all the hamiltonian
constraints $G_{\mu}$ $(\mu = 1,2,...,m)$ are first-class constraints
\cite{dirac}. From (\ref{2.25}) we conclude that our system has no
secondary constraints \cite{dirac}.

The identities (\ref{2.24}) are the first-order relations of the gauge
algebra in the hamiltonian formalism \cite{fradkin, henneaux}.
The constraints $G_{\mu}$ are also called the zeroth-order hamiltonian
structure functions \cite{fradkin, henneaux}. The first-order hamiltonian
structure functions are the functions $C_{\mu\nu}^{\alpha}$ in
(\ref{2.24}) \cite{fradkin, henneaux}.  

The second-order relations of the hamiltonian gauge algebra follow from the
Jacobi identities:

\begin{equation}
\{\{G_{\alpha}, G_{\beta}\}, G_{\gamma}\}
+ \{\{G_{\beta}, G_{\gamma}\}, G_{\alpha}\}
+ \{\{G_{\gamma}, G_{\alpha}\}, G_{\beta}\}
\equiv 0
\label{2.26}
\end{equation}

Indeed, from (\ref{2.26}) and (\ref{2.24}) it follows that:

\begin{equation}
\left(\{C_{\alpha\beta}^{\eta}, G_{\gamma}\} 
+ \{C_{\beta\gamma}^{\eta}, G_{\alpha}\} 
+ \{C_{\gamma\alpha}^{\eta}, G_{\beta}\} 
- C_{\alpha\beta}^{\delta} C_{\gamma\delta}^{\eta} 
- C_{\beta\gamma}^{\delta} C_{\alpha\delta}^{\eta} 
- C_{\gamma\alpha}^{\delta} C_{\beta\delta}^{\eta} 
\right) G_{\eta} \equiv 0
\label{2.27}
\end{equation}

From the irreducibility of the constraints (\ref{2.2}) it follows
\cite{fradkin} that there must exist some functions
$J_{\alpha\beta\gamma}^{\eta\sigma}$ in the momentum phase space $T^*Q$
such that:

\begin{equation}
\{C_{\alpha\beta}^{\eta}, G_{\gamma}\} 
+ \{C_{\beta\gamma}^{\eta}, G_{\alpha}\} 
+ \{C_{\gamma\alpha}^{\eta}, G_{\beta}\} 
- C_{\alpha\beta}^{\delta} C_{\gamma\delta}^{\eta} 
- C_{\beta\gamma}^{\delta} C_{\alpha\delta}^{\eta} 
- C_{\gamma\alpha}^{\delta} C_{\beta\delta}^{\eta} 
\equiv J_{\alpha\beta\gamma}^{\eta\sigma} G_{\sigma}
\label{2.28}
\end{equation}

\noindent or equivalently:

\begin{equation}
{\rm FL}^* \left(\{C_{\alpha\beta}^{\eta}, G_{\gamma}\} 
+ \{C_{\beta\gamma}^{\eta}, G_{\alpha}\} 
+ \{C_{\gamma\alpha}^{\eta}, G_{\beta}\} 
- C_{\alpha\beta}^{\delta} C_{\gamma\delta}^{\eta} 
- C_{\beta\gamma}^{\delta} C_{\alpha\delta}^{\eta} 
- C_{\gamma\alpha}^{\delta} C_{\beta\delta}^{\eta}\right)
\equiv 0 
\label{2.29}
\end{equation}

The identities (\ref{2.28}) are the so-called second-order relations of the
gauge algebra in the hamiltonian formalism \cite{fradkin,henneaux}. The
functions $J_{\alpha\beta\gamma}^{\eta\sigma}$ are the second-order
hamiltonian structure functions.

Our purpose in this paper is to write the lagrangian gauge structure
functions ($R_{\mu}^{i}$, $T_{\alpha\beta}^{\gamma}$,
$E_{\alpha\beta}^{ij}$, $D_{\alpha\beta\gamma}^{i\rho}$ and
$M_{\alpha\beta\gamma}^{ijk}$) explicitly in terms of the hamiltonian
structure functions $G_{\mu}$, $C_{\alpha\beta}^{\gamma}$ and their
derivatives. As we shall see, knowledge of the functional dependence of
$J_{\alpha\beta\gamma}^{\eta\sigma}$ or other higher order hamiltonian
structure functions is not required for this. 

Equations (\ref{2.10}) give us the lagrangian gauge generators
$R_{\mu}^{i}$. 

From (\ref{2.10}), (\ref{2.1}) and (\ref{1.3}) it immediately follows
that:

\begin{equation}
\frac{\partial R_{\mu}^{i}}{\partial\dot{q}^{k}} 
\equiv W_{kl} 
{\rm FL}^* \frac{\partial^{2} G_{\mu}}{\partial p_{l}\partial p_{i}}
\label{2.30}
\end{equation}

\begin{equation}
\frac{\partial R_{\mu}^{i}}{\partial q^{k}} 
\equiv 
{\rm FL}^* \frac{\partial^{2} G_{\mu}}{\partial q^{k} \partial p_{i}} 
+ \frac{\partial^{2} L}{\partial q^{k} \partial\dot{q}^{l}} 
{\rm FL}^* \frac{\partial^{2} G_{\mu}}{\partial p_{l} \partial p_{i}}
\label{2.31}
\end{equation}

Substituting (\ref{2.30}, \ref{2.31}) into (\ref{1.23}-\ref{1.25}) we
find:

\begin{eqnarray}
{\rm FL}^* \left(
\frac{\partial^{2} G_{\mu}}{\partial p_{i} \partial q^{j}} 
\frac{\partial G_{\nu}}{\partial p_{j}} 
- \frac{\partial^{2} G_{\nu}}{\partial p_{i} \partial q^{j}} 
\frac{\partial G_{\mu}}{\partial p_{j}} \right) 
+ {\rm FL}^* \left(\frac{\partial^{2} G_{\mu}}{\partial p_{i} \partial
p_{l}} \frac{\partial G_{\nu}}{\partial p_{j}} 
- \frac{\partial^{2} G_{\nu}}{\partial p_{i} \partial p_{l}} 
\frac{\partial G_{\mu}}{\partial p_{j}}\right) 
\frac{\partial^{2} L}{\partial\dot{q}^{l}\partial q^{j}} + \nonumber & &\\
+ {\rm FL}^* \left(
\frac{\partial^{2} G_{\mu}}{\partial p_{i} \partial p_{l}} 
\frac{\partial^{2} G_{\nu}}{\partial p_{m} \partial p_{j}}
- \frac{\partial^{2} G_{\nu}}{\partial p_{i} \partial p_{l}} 
\frac{\partial^{2} G_{\mu}}{\partial p_{m} \partial p_{j}}\right) 
W_{lm} \frac{\partial^{2} L}{\partial\dot{q}^{j} \partial q^{k}}
\dot{q}^{k} + \nonumber & &\\
+ {\rm FL}^* \left(
\frac{\partial^{2} G_{\mu}}{\partial p_{i} \partial p_{l}} 
\frac{\partial^{2} G_{\nu}}{\partial p_{m} \partial q^{k}}
- \frac{\partial^{2} G_{\nu}}{\partial p_{i} \partial p_{l}} 
\frac{\partial^{2} G_{\mu}}{\partial p_{m} \partial q^{k}}\right) 
\dot{q}^{k} W_{lm} \equiv 
T_{\mu\nu}^{\gamma} R_{\gamma}^{i} - E_{\mu\nu}^{ij} \alpha_{j}
\label{2.32}
\end{eqnarray}

\begin{equation}
\left(
{\rm FL}^* \frac{\partial^{2} G_{\mu}}{\partial p_{i} \partial p_{l}}
W_{lm} 
{\rm FL}^* \frac{\partial^{2} G_{\nu}}{\partial p_{m} \partial p_{j}}
- {\rm FL}^* \frac{\partial^{2} G_{\nu}}{\partial p_{i} \partial p_{l}}
W_{lm} 
{\rm FL}^* \frac{\partial^{2} G_{\mu}}{\partial p_{m} \partial p_{j}}
\right) W_{jk} \equiv E_{\mu\nu}^{ij} W_{jk}
\label{2.33}
\end{equation}

\begin{equation}
{\rm FL}^* \frac{\partial^{2} G_{\mu}}{\partial p_{i} \partial p_{k}}
W_{kj} R_{\nu}^{j} \equiv 0
\label{2.34}
\end{equation}

Substituting (\ref{2.16}) and (\ref{2.10}, \ref{2.11}) into (\ref{2.32})
and using (\ref{1.4}) we see that the identities (\ref{2.32}) can be
rewritten as:

\begin{eqnarray}
{\rm FL}^* \frac{\partial}{\partial p_{i}} \{G_{\mu}, G_{\nu}\}
- \left(
{\rm FL}^* \frac{\partial^{2} G_{\mu}}{\partial p_{i} \partial p_{l}}
W_{lm} 
{\rm FL}^* \frac{\partial^{2} G_{\nu}}{\partial p_{m} \partial p_{j}}
- {\rm FL}^* \frac{\partial^{2} G_{\nu}}{\partial p_{i} \partial p_{l}}
W_{lm} 
{\rm FL}^* \frac{\partial^{2} G_{\mu}}{\partial p_{m} \partial p_{j}}
\right) \alpha_{j} \nonumber\\
+ {\rm FL}^* \frac{\partial^{2} G_{\mu}}{\partial p_{i} \partial_{l}}
\left(B_{lj}{\rm FL}^* \frac{\partial G_{\nu}}{\partial p_{j}}
+ W_{lm}{\rm FL}^* \frac{\partial^{2} G_{\nu}}{\partial p_{m} \partial
p_{j}} \frac{\partial L}{\partial q^{j}} 
+ W_{lm} {\rm FL}^* \frac{\partial^{2} G_{\nu}}{\partial p_{m} \partial
q^{k}}\dot{q}^{k}\right) \nonumber \\
- {\rm FL}^* \frac{\partial^{2} G_{\nu}}{\partial p_{i} \partial_{l}}
\left(B_{lj}{\rm FL}^* \frac{\partial G_{\mu}}{\partial p_{j}}
+ W_{lm}{\rm FL}^* \frac{\partial^{2} G_{\mu}}{\partial p_{m} \partial
p_{j}} \frac{\partial L}{\partial q^{j}} 
+ W_{lm} {\rm FL}^* \frac{\partial^{2} G_{\mu}}{\partial p_{m} \partial
q^{k}}\dot{q}^{k}\right) \nonumber \\
\equiv T_{\mu\nu}^{\gamma} R_{\gamma}^{i} - E_{\mu\nu}^{ij} \alpha_{j}
\label{2.37}
\end{eqnarray}

Notice that:

\begin{equation}
{\rm FL}^* \frac{\partial}{\partial p_{i}} \{G_{\mu}, G_{\nu}\}
\equiv {\rm FL}^* \left(\frac{\partial C_{\mu\nu}^{\gamma}}{\partial
p_{i}} G_{\gamma} + C_{\mu\nu}^{\gamma} \frac{\partial
G_{\gamma}}{\partial p_{i}} \right)
\equiv {\rm FL}^* C_{\mu\nu}^{\gamma} 
{\rm FL}^* \frac{\partial G_{\gamma}}{\partial p_{i}}
\label{2.39}
\end{equation}

On the other hand, from (\ref{2.20}), (\ref{2.30}) and (\ref{2.31}) it
follows that: 

\begin{equation}
B_{lj}{\rm FL}^* \frac{\partial G_{\alpha}}{\partial p_{j}}
+ W_{lm}{\rm FL}^* \frac{\partial^{2} G_{\alpha}}{\partial p_{m} \partial
p_{j}} \frac{\partial L}{\partial q^{j}} 
+ W_{lm} {\rm FL}^* \frac{\partial^{2} G_{\alpha}}{\partial p_{m} \partial
q^{k}}\dot{q}^{k} \equiv 0
\label{2.40}
\end{equation}

Finally, substituting (\ref{2.39}) and (\ref{2.40}) into (\ref{2.37}) we
obtain the identities:

\begin{eqnarray}
{\rm FL}^* C_{\mu\nu}^{\gamma} R_{\gamma}^{i} 
- \left(
{\rm FL}^* \frac{\partial^{2} G_{\mu}}{\partial p_{i} \partial p_{l}}
W_{lm} 
{\rm FL}^* \frac{\partial^{2} G_{\nu}}{\partial p_{m} \partial p_{j}}
- {\rm FL}^* \frac{\partial^{2} G_{\nu}}{\partial p_{i} \partial p_{l}}
W_{lm} 
{\rm FL}^* \frac{\partial^{2} G_{\mu}}{\partial p_{m} \partial p_{j}}
\right) \alpha_{j} \nonumber\\
\equiv T_{\mu\nu}^{\gamma} R_{\gamma}^{i} - E_{\mu\nu}^{ij} \alpha_{j}
\label{2.41}
\end{eqnarray}

From the identities (\ref{2.41}) and (\ref{2.33}) it follows that we can
make the following identification:

\begin{equation}
T_{\mu\nu}^{\gamma} = {\rm FL}^* C_{\mu\nu}^{\gamma}
\label{2.42}
\end{equation}

\begin{equation}
E_{\mu\nu}^{ij} = 
{\rm FL}^* \frac{\partial^{2} G_{\mu}}{\partial p_{i} \partial p_{l}}
W_{lm} 
{\rm FL}^* \frac{\partial^{2} G_{\nu}}{\partial p_{m} \partial p_{j}}
- {\rm FL}^* \frac{\partial^{2} G_{\nu}}{\partial p_{i} \partial p_{l}}
W_{lm} 
{\rm FL}^* \frac{\partial^{2} G_{\mu}}{\partial p_{m} \partial p_{j}}
\label{2.43}
\end{equation}

Equations (\ref{2.42}) and (\ref{2.43}) give us the lagrangian
second-order gauge structure
functions in terms of the hamiltonian first-order structure functions and
the second derivatives of the hamiltonian constraints with respect to the
canonical momenta.

Our task now is to find an expression for the third-order lagrangian 
gauge structure functions $D_{\alpha\beta\gamma}^{i\rho}$. From
(\ref{2.42}) it follows that:

\begin{equation}
\frac{\partial T_{\alpha\beta}^{\rho}}{\partial q^{j}} 
= {\rm FL}^* \frac{\partial C_{\alpha\beta}^{\rho}}{\partial q^{j}} +
\frac{\partial^{2} L}{\partial q^{j} \partial\dot{q}^{k}} 
{\rm FL}^* \frac{\partial C_{\alpha\beta}^{\rho}}{\partial p_{k}}
\label{2.44}
\end{equation}

\begin{equation}
\frac{\partial T_{\alpha\beta}^{\rho}}{\partial\dot{q}^{j}} 
= W_{jk} {\rm FL}^* \frac{\partial C_{\alpha\beta}^{\rho}}{\partial p_{k}}
\label{2.45}
\end{equation}

Notice also that using (\ref{2.30}, \ref{2.31}) and (\ref{1.2}),
(\ref{1.4}) we can write:

\begin{equation}
\dot{R}_{\alpha}^{j} 
= {\rm FL}^* \frac{\partial^{2} G_{\alpha}}{\partial p_{j} \partial q^{l}}
\dot{q}^{l}
+ {\rm FL}^* \frac{\partial^{2} G_{\alpha}}{\partial p_{j} \partial p_{l}}
\frac{\partial L}{\partial q^{l}}
+ {\rm FL}^* \frac{\partial^{2} G_{\alpha}}{\partial p_{j} \partial p_{k}}
L_{k} 
\label{2.47}
\end{equation}

Substituting (\ref{2.42}), (\ref{2.44}), (\ref{2.45}) and (\ref{2.47}) into
(\ref{1.38}) we obtain:

\begin{eqnarray}
A_{\alpha\beta\gamma}^{\delta} =
{1 \over 3} \left[
{\rm FL}^* \left(C_{\alpha\eta}^{\rho} C_{\beta\gamma}^{\eta} 
+ C_{\beta\eta}^{\rho} C_{\gamma\alpha}^{\eta} 
+ C_{\gamma\eta}^{\rho} C_{\alpha\beta}^{\eta}\right) 
- {\rm FL}^* \left(\frac{\partial G_{\alpha}}{\partial p_{j}} 
\frac{\partial C_{\beta\gamma}^{\rho}}{\partial q^{j}}
+ \frac{\partial G_{\beta}}{\partial p_{j}} 
\frac{\partial C_{\gamma\alpha}^{\rho}}{\partial q^{j}}
+ \frac{\partial G_{\gamma}}{\partial p_{j}} 
\frac{\partial C_{\alpha\beta}^{\rho}}{\partial q^{j}}
\right)\right. \nonumber \\
- \frac{\partial^{2} L}{\partial q^{j} \partial\dot{q}^{k}}
{\rm FL}^* \left(\frac{\partial G_{\alpha}}{\partial p_{j}} 
\frac{\partial C_{\beta\gamma}^{\rho}}{\partial p_{k}}
+ \frac{\partial G_{\beta}}{\partial p_{j}} 
\frac{\partial C_{\gamma\alpha}^{\rho}}{\partial p_{k}}
+ \frac{\partial G_{\gamma}}{\partial p_{j}} 
\frac{\partial C_{\alpha\beta}^{\rho}}{\partial p_{k}}
\right) \nonumber \\
- \dot{q}^{l} 
\left(
{\rm FL}^* \frac{\partial^{2} G_{\alpha}}{\partial q^{l} \partial p_{j}}
W_{jk} 
{\rm FL}^* \frac{\partial C_{\beta\gamma}^{\rho}}{\partial p_{k}}
+ {\rm FL}^* \frac{\partial^{2} G_{\beta}}{\partial q^{l} \partial p_{j}}
W_{jk} 
{\rm FL}^* \frac{\partial C_{\gamma\alpha}^{\rho}}{\partial p_{k}}
+ {\rm FL}^* \frac{\partial^{2} G_{\gamma}}{\partial q^{l} \partial p_{j}}
W_{jk} 
{\rm FL}^* \frac{\partial C_{\alpha\beta}^{\rho}}{\partial p_{k}}
\right) \nonumber \\
- \frac{\partial L}{\partial q^{l}}
\left({\rm FL}^* \frac{\partial^{2} G_{\alpha}}{\partial p_{l} \partial
p_{j}}
W_{jk} 
{\rm FL}^* \frac{\partial C_{\beta\gamma}^{\rho}}{\partial p_{k}}
+ {\rm FL}^* \frac{\partial^{2} G_{\beta}}{\partial p_{l} \partial p_{j}}
W_{jk} 
{\rm FL}^* \frac{\partial C_{\gamma\alpha}^{\rho}}{\partial p_{k}}
+ {\rm FL}^* \frac{\partial^{2} G_{\gamma}}{\partial p_{l} \partial p_{j}}
W_{jk} 
{\rm FL}^* \frac{\partial C_{\alpha\beta}^{\rho}}{\partial p_{k}}
\right) \nonumber \\
\left.- L_{j} \left({\rm FL}^* \frac{\partial^{2} G_{\alpha}}{\partial
p_{j}
\partial p_{l}}
W_{lk} 
{\rm FL}^* \frac{\partial C_{\beta\gamma}^{\rho}}{\partial p_{k}}
+ {\rm FL}^* \frac{\partial^{2} G_{\beta}}{\partial p_{j} \partial p_{l}}
W_{lk} 
{\rm FL}^* \frac{\partial C_{\gamma\alpha}^{\rho}}{\partial p_{k}}
+ {\rm FL}^* \frac{\partial^{2} G_{\gamma}}{\partial p_{j} \partial p_{l}}
W_{lk} 
{\rm FL}^* \frac{\partial C_{\alpha\beta}^{\rho}}{\partial p_{k}}
\right) \right] \nonumber \\
\label{2.48}
\end{eqnarray}

Using (\ref{2.16}) and (\ref{2.10},\ref{2.11}) we can rewrite (\ref{2.48})
as follows:

\begin{eqnarray}
A_{\alpha\beta\gamma}^{\delta} =
{1 \over 3} \left[
{\rm FL}^* \left(- \{C_{\alpha\beta}^{\rho}, G_{\gamma}\}
- \{C_{\gamma\alpha}^{\rho}, G_{\beta}\}
- \{C_{\beta\gamma}^{\rho}, G_{\alpha}\}
+ C_{\alpha\eta}^{\rho} C_{\beta\gamma}^{\eta} 
+ C_{\beta\eta}^{\rho} C_{\gamma\alpha}^{\eta} 
+ C_{\gamma\eta}^{\rho} C_{\alpha\beta}^{\eta}\right)\right. \nonumber \\
- {\rm FL}^* \frac{\partial C_{\alpha\beta}^{\rho}}{\partial p_{k}} 
\left( B_{kj}{\rm FL}^* \frac{\partial G_{\gamma}}{\partial p_{j}}
+ W_{jk}{\rm FL}^* \frac{\partial^{2} G_{\gamma}}{\partial p_{l} \partial
p_{j}} \frac{\partial L}{\partial q^{l}} 
+ W_{jk} {\rm FL}^* \frac{\partial^{2} G_{\gamma}}{\partial p_{j} \partial
q^{l}}\dot{q}^{l} \right) \nonumber \\ 
- {\rm FL}^* \frac{\partial C_{\gamma\alpha}^{\rho}}{\partial p_{k}} 
\left( B_{kj}{\rm FL}^* \frac{\partial G_{\beta}}{\partial p_{j}}
+ W_{jk}{\rm FL}^* \frac{\partial^{2} G_{\beta}}{\partial p_{l} \partial
p_{j}} \frac{\partial L}{\partial q^{l}} 
+ W_{jk} {\rm FL}^* \frac{\partial^{2} G_{\beta}}{\partial p_{j} \partial
q^{l}}\dot{q}^{l} \right) \nonumber \\ 
- {\rm FL}^* \frac{\partial C_{\beta\gamma}^{\rho}}{\partial p_{k}} 
\left( B_{kj}{\rm FL}^* \frac{\partial G_{\alpha}}{\partial p_{j}}
+ W_{jk}{\rm FL}^* \frac{\partial^{2} G_{\alpha}}{\partial p_{l} \partial
p_{j}} \frac{\partial L}{\partial q^{l}} 
+ W_{jk} {\rm FL}^* \frac{\partial^{2} G_{\alpha}}{\partial p_{j} \partial
q^{l}}\dot{q}^{l} \right) \nonumber \\ 
\left.- L_{j} \left({\rm FL}^* \frac{\partial^{2} G_{\alpha}}{\partial
p_{j} \partial p_{l}}
W_{lk} 
{\rm FL}^* \frac{\partial C_{\beta\gamma}^{\rho}}{\partial p_{k}}
+ {\rm FL}^* \frac{\partial^{2} G_{\beta}}{\partial p_{j} \partial p_{l}}
W_{lk} 
{\rm FL}^* \frac{\partial C_{\gamma\alpha}^{\rho}}{\partial p_{k}}
+ {\rm FL}^* \frac{\partial^{2} G_{\gamma}}{\partial p_{j} \partial p_{l}}
W_{lk} 
{\rm FL}^* \frac{\partial C_{\alpha\beta}^{\rho}}{\partial p_{k}}
\right) \right] \nonumber \\
\label{2.49}
\end{eqnarray}

Finally, using (\ref{2.29}) and (\ref{2.40}) we obtain:

\begin{eqnarray}
A_{\alpha\beta\gamma}^{\delta} =
- {1 \over 3} L_{j} \left({\rm FL}^* \frac{\partial^{2}
G_{\alpha}}{\partial
p_{j} \partial p_{l}}
W_{lk} 
{\rm FL}^* \frac{\partial C_{\beta\gamma}^{\rho}}{\partial p_{k}}
+ {\rm FL}^* \frac{\partial^{2} G_{\beta}}{\partial p_{j} \partial p_{l}}
W_{lk} 
{\rm FL}^* \frac{\partial C_{\gamma\alpha}^{\rho}}{\partial p_{k}}
+ {\rm FL}^* \frac{\partial^{2} G_{\gamma}}{\partial p_{j} \partial p_{l}}
W_{lk} 
{\rm FL}^* \frac{\partial C_{\alpha\beta}^{\rho}}{\partial p_{k}}\right) 
\nonumber \\
\label{2.50}
\end{eqnarray}

From (\ref{1.41}) and (\ref{2.50}) we see that the third-order lagrangian
gauge structure functions $D_{\alpha\beta\gamma}^{i\rho}$ can be written
as:

\begin{eqnarray}
D_{\alpha\beta\gamma}^{i\rho} =
- {1 \over 3}\left({\rm FL}^* \frac{\partial^{2}
G_{\alpha}}{\partial
p_{i} \partial p_{j}}
W_{jk} 
{\rm FL}^* \frac{\partial C_{\beta\gamma}^{\rho}}{\partial p_{k}}
+ {\rm FL}^* \frac{\partial^{2} G_{\beta}}{\partial p_{i} \partial p_{j}}
W_{jk} 
{\rm FL}^* \frac{\partial C_{\gamma\alpha}^{\rho}}{\partial p_{k}}
+ {\rm FL}^* \frac{\partial^{2} G_{\gamma}}{\partial p_{i} \partial p_{j}}
W_{jk} 
{\rm FL}^* \frac{\partial C_{\alpha\beta}^{\rho}}{\partial p_{k}}\right) 
\nonumber \\
\label{2.51}
\end{eqnarray}

The derivation presented above can also be viewed as a proof by
construction of the existence of the gauge structure functions
$D_{\alpha\beta\gamma}^{i\rho}$. 

Using (\ref{1.6}), (\ref{2.30}), (\ref{2.43}) and (\ref{2.45}) one can
easily prove the identities (\ref{1.35}).

Let us now find the fourth-order lagrangian gauge structure functions
$M_{\alpha\beta\gamma}^{ijk}$. For that, we need to write the
left-hand-side of (\ref{1.45}) in terms of hamiltonian quantities.
After some lengthy calculations we find:

\begin{eqnarray}
R_{\rho}^{i} D_{\alpha\beta\gamma}^{j\rho} 
- R_{\rho}^{j} D_{\alpha\beta\gamma}^{i\rho} &\equiv& \nonumber \\
& &- {1 \over 3} \left[ 
E_{\alpha\eta}^{ij} T_{\beta\gamma}^{\eta} 
+ E_{\beta\eta}^{ij} T_{\gamma\alpha}^{\eta} 
+ E_{\gamma\eta}^{ij} T_{\alpha\beta}^{\eta} \right. \nonumber \\
& & + {\rm FL}^* \left(
\frac{\partial^{2}}{\partial p_{i} \partial p_{k}}\{G_{\alpha},G_{\beta}\}
\frac{\partial^{2} G_{\gamma}}{\partial p_{l} \partial p_{j}}
-\frac{\partial^{2}}{\partial p_{j} \partial
p_{k}}\{G_{\alpha},G_{\beta}\}
\frac{\partial^{2} G_{\gamma}}{\partial p_{l} \partial p_{i}}\right)
W_{kl}
\nonumber \\ 
& &+ {\rm FL}^* \left(
\frac{\partial^{2}}{\partial p_{i} \partial p_{k}}\{G_{\beta},G_{\gamma}\}
\frac{\partial^{2} G_{\alpha}}{\partial p_{l} \partial p_{j}}
-\frac{\partial^{2}}{\partial p_{j} \partial
p_{k}}\{G_{\beta},G_{\gamma}\}
\frac{\partial^{2} G_{\alpha}}{\partial p_{l} \partial p_{i}}\right)
W_{kl} \nonumber \\
& &\left. + {\rm FL}^* \left(
\frac{\partial^{2}}{\partial p_{i} \partial
p_{k}}\{G_{\gamma},G_{\alpha}\}
\frac{\partial^{2} G_{\beta}}{\partial p_{l} \partial p_{j}}
-\frac{\partial^{2}}{\partial p_{j} \partial
p_{k}}\{G_{\gamma},G_{\alpha}\}
\frac{\partial^{2} G_{\beta}}{\partial p_{l} \partial p_{j}}\right) W_{kl} 
\right] \nonumber \\
\label{2.52}
\end{eqnarray}

\begin{eqnarray}
B_{\alpha\beta\gamma}^{ij} &\equiv& 
{1 \over 3} \left[ 
E_{\alpha\eta}^{ij} T_{\beta\gamma}^{\eta} 
+ E_{\beta\eta}^{ij} T_{\gamma\alpha}^{\eta} 
+ E_{\gamma\eta}^{ij} T_{\alpha\beta}^{\eta} \right. \nonumber \\
& & + {\rm FL}^* \left(
\frac{\partial^{2}}{\partial p_{i} \partial p_{k}}\{G_{\alpha},G_{\beta}\}
\frac{\partial^{2} G_{\gamma}}{\partial p_{l} \partial p_{j}}
-\frac{\partial^{2}}{\partial p_{j} \partial
p_{k}}\{G_{\alpha},G_{\beta}\}
\frac{\partial^{2} G_{\gamma}}{\partial p_{l} \partial p_{i}}\right)
W_{kl}\nonumber \\ 
& &+ {\rm FL}^* \left(
\frac{\partial^{2}}{\partial p_{i} \partial p_{k}}\{G_{\beta},G_{\gamma}\}
\frac{\partial^{2} G_{\alpha}}{\partial p_{l} \partial p_{j}}
-\frac{\partial^{2}}{\partial p_{j} \partial
p_{k}}\{G_{\beta},G_{\gamma}\}
\frac{\partial^{2} G_{\alpha}}{\partial p_{l} \partial p_{i}}\right)
W_{kl} \nonumber \\
& & \left. + {\rm FL}^* \left(
\frac{\partial^{2}}{\partial p_{i} \partial
p_{k}}\{G_{\gamma},G_{\alpha}\}
\frac{\partial^{2} G_{\beta}}{\partial p_{l} \partial p_{j}}
-\frac{\partial^{2}}{\partial p_{j} \partial
p_{k}}\{G_{\gamma},G_{\alpha}\}
\frac{\partial^{2} G_{\beta}}{\partial p_{l} \partial p_{j}}\right) W_{kl} 
\right] \nonumber \\
& & - {1 \over 3} L_{k} \left[
{\rm FL}^* \frac{\partial^{2} G_{\alpha}}{\partial p_{k} \partial p_{l}}
\frac{\partial E_{\beta\gamma}^{ij}}{\partial\dot{q}^{l} }
+ {\rm FL}^* \frac{\partial^{2} G_{\beta}}{\partial p_{k} \partial p_{l}}
\frac{\partial E_{\gamma\alpha}^{ij}}{\partial\dot{q}^{l} }
+ {\rm FL}^* \frac{\partial^{2} G_{\gamma}}{\partial p_{k} \partial p_{l}}
\frac{\partial E_{\alpha\beta}^{ij}}{\partial\dot{q}^{l} }\right. 
\nonumber \\
& & + {\rm FL}^* \frac{\partial^{3} G_{\alpha}}{\partial p_{k} \partial
p_{m}
\partial p_{n}} 
\left(\frac{\partial R_{\beta}^{i}}{\partial\dot{q}^{m}} 
\frac{\partial R_{\gamma}^{j}}{\partial\dot{q}^{n} } 
- \frac{\partial R_{\gamma}^{i}}{\partial\dot{q}^{m}} 
\frac{\partial R_{\beta}^{j}}{\partial\dot{q}^{n} } \right) 
\nonumber \\
& & + {\rm FL}^* \frac{\partial^{3} G_{\beta}}{\partial p_{k} \partial
p_{m}
\partial p_{n}} 
\left(\frac{\partial R_{\gamma}^{i}}{\partial\dot{q}^{m}} 
\frac{\partial R_{\alpha}^{j}}{\partial\dot{q}^{n} } 
- \frac{\partial R_{\alpha}^{i}}{\partial\dot{q}^{m}} 
\frac{\partial R_{\gamma}^{j}}{\partial\dot{q}^{n} } \right) \nonumber \\
& & \left. + {\rm FL}^* \frac{\partial^{3} G_{\gamma}}{\partial p_{k}
\partial p_{m} \partial p_{n}} 
\left(\frac{\partial R_{\alpha}^{i}}{\partial\dot{q}^{m}} 
\frac{\partial R_{\beta}^{j}}{\partial\dot{q}^{n} } 
- \frac{\partial R_{\beta}^{i}}{\partial\dot{q}^{m}} 
\frac{\partial R_{\alpha}^{j}}{\partial\dot{q}^{n} } \right) \right]
\label{2.53}
\end{eqnarray}

From (\ref{1.45}) and (\ref{2.52}, \ref{2.53}) we see that it is possible
to make the following identification:

\begin{eqnarray}
M_{\alpha\beta\gamma}^{ijk} &\equiv&
- {1 \over 3} \left[
{\rm FL}^* \frac{\partial^{2} G_{\alpha}}{\partial p_{k} \partial p_{l}}
\frac{\partial E_{\beta\gamma}^{ij}}{\partial\dot{q}^{l} }
+ {\rm FL}^* \frac{\partial^{2} G_{\beta}}{\partial p_{k} \partial p_{l}}
\frac{\partial E_{\gamma\alpha}^{ij}}{\partial\dot{q}^{l} }
+ {\rm FL}^* \frac{\partial^{2} G_{\gamma}}{\partial p_{k} \partial p_{l}}
\frac{\partial E_{\alpha\beta}^{ij}}{\partial\dot{q}^{l} }\right. 
\nonumber \\
& & + {\rm FL}^* \frac{\partial^{3} G_{\alpha}}{\partial p_{k} \partial
p_{m}
\partial p_{n}} 
\left(\frac{\partial R_{\beta}^{i}}{\partial\dot{q}^{m}} 
\frac{\partial R_{\gamma}^{j}}{\partial\dot{q}^{n} } 
- \frac{\partial R_{\gamma}^{i}}{\partial\dot{q}^{m}} 
\frac{\partial R_{\beta}^{j}}{\partial\dot{q}^{n} } \right) 
\nonumber \\
& & + {\rm FL}^* \frac{\partial^{3} G_{\beta}}{\partial p_{k} \partial
p_{m}
\partial p_{n}} 
\left(\frac{\partial R_{\gamma}^{i}}{\partial\dot{q}^{m}} 
\frac{\partial R_{\alpha}^{j}}{\partial\dot{q}^{n} } 
- \frac{\partial R_{\alpha}^{i}}{\partial\dot{q}^{m}} 
\frac{\partial R_{\gamma}^{j}}{\partial\dot{q}^{n} } \right) \nonumber \\
& & \left. + {\rm FL}^* \frac{\partial^{3} G_{\gamma}}{\partial p_{k}
\partial p_{m} \partial p_{n}} 
\left(\frac{\partial R_{\alpha}^{i}}{\partial\dot{q}^{m}} 
\frac{\partial R_{\beta}^{j}}{\partial\dot{q}^{n} } 
- \frac{\partial R_{\beta}^{i}}{\partial\dot{q}^{m}} 
\frac{\partial R_{\alpha}^{j}}{\partial\dot{q}^{n} } \right) \right]
\label{2.54}
\end{eqnarray}

Using (\ref{2.30}) and (\ref{2.43}) we can rewrite (\ref{2.54}) in terms
of the hamiltonian constraints as follows:

\begin{eqnarray}
M_{\alpha\beta\gamma}^{ijk} &\equiv&
- {1 \over 3} \left[
{\rm FL}^* \frac{\partial^{2} G_{\alpha}}{\partial p_{k} \partial p_{l}}
{\it P}_{l \beta\gamma}^{ij}
+ {\rm FL}^* \frac{\partial^{2} G_{\beta}}{\partial p_{k} \partial p_{l}}
{\it P}_{l \gamma\alpha}^{ij}
+ {\rm FL}^* \frac{\partial^{2} G_{\gamma}}{\partial p_{k} \partial p_{l}}
{\it P}_{l \alpha\beta}^{ij}\right. 
\nonumber \\
& & + {\rm FL}^* \frac{\partial^{3} G_{\alpha}}{\partial p_{k} \partial
p_{m}
\partial p_{n}} 
\left({\it P}_{m\beta}^{i} {\it P}_{n\gamma}^{j} 
- {\it P}_{m\gamma}^{i} {\it P}_{n\beta}^{j}\right) 
\nonumber \\
& & + {\rm FL}^* \frac{\partial^{3} G_{\beta}}{\partial p_{k} \partial
p_{m}
\partial p_{n}} 
\left({\it P}_{m\gamma}^{i} {\it P}_{n\alpha}^{j} 
- {\it P}_{m\alpha}^{i} {\it P}_{n\gamma}^{j}\right) 
\nonumber \\
& & \left. + {\rm FL}^* \frac{\partial^{3} G_{\gamma}}{\partial p_{k}
\partial p_{m} \partial p_{n}} 
\left({\it P}_{m\alpha}^{i} {\it P}_{n\beta}^{j} 
- {\it P}_{m\beta}^{i} {\it P}_{n\alpha}^{j}\right) 
\right]
\label{2.55}
\end{eqnarray}

\noindent where,

\begin{equation}
{\it P}_{j\mu}^{i} = 
W_{jk} {\rm FL}^* \frac{\partial^{2} G_{\mu}}{\partial p_{k} \partial
p_{i}}
\label{2.56}
\end{equation}

\begin{eqnarray}
{\it P}_{k\mu\nu}^{ij} & = & 
\frac{\partial W_{lm}}{\partial\dot{q}^{k}} 
{\rm FL}^* \left(
\frac{\partial^{2} G_{\mu}}{\partial p_{i} \partial p_{l}} 
\frac{\partial^{2} G_{\nu}}{\partial p_{m} \partial p_{j}} 
- \frac{\partial^{2} G_{\nu}}{\partial p_{i} \partial p_{l}} 
\frac{\partial^{2} G_{\mu}}{\partial p_{m} \partial p_{j}} 
\right)\nonumber \\
& & + W_{lm}W_{kn} {\rm FL}^* \frac{\partial}{\partial p_{n}} 
\left(
\frac{\partial^{2} G_{\mu}}{\partial p_{i} \partial p_{l}} 
\frac{\partial^{2} G_{\nu}}{\partial p_{m} \partial p_{j}} 
- \frac{\partial^{2} G_{\nu}}{\partial p_{i} \partial p_{l}} 
\frac{\partial^{2} G_{\mu}}{\partial p_{m} \partial p_{j}} 
\right)
\label{2.57}
\end{eqnarray}

\vskip 1cm

The main results presented in this paper are the explicit expressions for
the lagrangian gauge structure functions  
$E^{ij}_{\mu\nu}$,
$D_{\alpha\beta\gamma}^{i\rho}$ and
$M_{\alpha\beta\gamma}^{ijk}$ given by the equations
(\ref{2.43}), (\ref{2.51}) and (\ref{2.55}). 
These equations  show  how the higher-order lagrangian structure tensors
are determined
by the hamiltonian constraints and the hamiltonian
first-order structure functions. To determine these
lagrangian structure tensors no knowledge of the higher-order
hamiltonian structure functions is required. 
Notice that for the lagrangian gauge algebra to be open
($E_{\mu\nu}^{ij} \neq 0$) it is necessary to have at least two
constraints that depend nonlinearly on the momenta. In order to have 
nonvanishing third-order structure functions
$D_{\alpha\beta\gamma}^{i\rho}$,
the first-order structure functions $C_{\mu\nu}^{\eta}$ must depend on the
canonical momenta $p_{i}$. The third derivatives of the constraints with
respect to the canonical momenta determine the fourth-order structure
tensors $M_{\alpha\beta\gamma}^{ijk}$. The method presented here can be
used to obtain tensors of even higher orders.  

\vskip 1cm
I would like to thank Gordon Semenoff for helpful discussions and
comments.

\end{document}